\documentclass[10pt]{article}
\usepackage{amsmath,amssymb,amsfonts}
\usepackage{units}
\usepackage{bbold,euler}
\usepackage{graphicx}
\usepackage{dcolumn}
\usepackage{bm,bbm,mathtools,esvect}
\usepackage{slashed}
\usepackage{esvect}
\usepackage{fullpage}
\usepackage[utf8]{inputenc}
\title{\bf QUATERNIONIC FERMIONIC FIELD }
\author{{\sf SERGIO GIARDINO\footnote{\tt sergio.giardino@ufrgs.br}}\\
\\
\small \it Departamento de Matem\'atica Pura e Aplicada \\
\small \it Universidade Federal do Rio Grande do Sul (UFRGS)\\
\small \it Brazil}
\begin{document}
\date{} 
\maketitle

\begin{abstract}
\noindent The second quantization of the quaternionic fermionic field is undertaken using the
real Hilbert space approach to quaternionic quantum mechanics ($\mathbbm H$QM). The solution
 responds to an open problem of quaternionic quantum theory, and launches the basis to the development
of the quaternionic interaction theory. 
%
%
%
%
%
\end{abstract}

\section{ Introduction\label{I}}

The introduction of mathematical structure is an essential requirement in order to improve a physical theory. The more sophisticated the mathematical framework, the more subtle the physical processes the theory may describe. Although seemingly simple, this method for generalizing physical theories has the risk of unnecessary complication if the mathematical theory is too involved, and therefore suitable criteria to select the mathematical framework are essential in order to obtain meaningful physical theories. 

As an example, we remember that quaternions ($\mathbbm H$) are generalized complex numbers \cite{Morais:2014rqc,Garling:2011zz,Ward:1997qcn} that seem to be the simplest mathematical artifact in order to generalize the physical theories formulated in terms of complex numbers ($\mathbbm C$), namely quantum mechanics and quantum field theory.
The  first implementation of quantum mechanics within a quaternionic framework took benefit of anti-hermitian energy operators \cite{Adler:1995qqm}, and found serious hindrances, such as  the breakdown of the Ehrenfest theorem ({\it cf.} Section 4.4 of \cite{Adler:1995qqm}).
However, the formulation of  quaternionic quantum mechanics ($\mathbbm H$QM) in terms of a real Hilbert space  \cite{Giardino:2018lem,Giardino:2018rhs} solved these conceptual difficulties, and many novel solutions have been obtained 
\cite{Giardino:2016xap,Giardino:2017yke,Giardino:2017pqq,Giardino:2019xwm,Giardino:2020ztf,Giardino:2020cee,Giardino:2021ofo}. 

The inconsistencies of anti-hermitian non-relativistic $\mathbbm H$QM are also found in the relativistic theory. However, and despite them, several attempts have been done in order to investigate whether a relativistic  quaternionic quantum theory could be obtained. In terms of the scalar field, we quote \cite{Horwitz:1984ep,Adler:1985uh,Nash:1985xf,Nash:1987sj,DeLeo:1991mi,DeLeo:1996ac,Giardino:2012ti,Giardino:2015ola,Giardino:2015dza,Steinberg:2020xvf} as former attempts using the anti-hermitian formalism. On the other hand, the real Hilbert approach enabled us to obtain the solutions of the Klein-Gordon equation \cite{Giardino:2021lov}, of the Dirac equation \cite{Giardino:2021mjj}, as well as the second quantization of the scalar field  within a quaternionic formalism \cite{Giardino:2022kxk} for the first time.

In this article,  the  method previously developed for  quantizing the quaternionic scalar field \cite{Giardino:2022kxk} will be used to quantize the quaternionic fermionic field, and this solution is absolutely new. Although several solutions to the Dirac equation using quaternionic the anti-hermitian scheme are already known
\cite{Rotelli:1988fc,Adler:1989xf,Davies:1990pm,DeLeo:1995yq,Gursey:1996mj,DeLeo:1996ger,Rawat:2007vm,DeLeo:2013xfa,DeLeo:2015hza,Giardino:2015iia,Kober:2015bkv,Hassanabadi:2017jiz,Hassanabadi:2017wrt}, the quantization of a quaternionic fermionic field was never obtained within an anti-hermitian program, with the caveat of the applications of quaternions to the Dirac equation within the usual framework of relativistic $\mathbbm C$QM \cite{Gursey:1955cbd,Gursey:1957rkc,Murray:1991cad}. However, this quaternionic approach to complex relativistic quantum mechanics lies outside the scope of $\mathbbm H$QM, and outside of the approach of this article as well. 

Our results demonstrate that the  quantization in terms of complex components is suitable to  the Dirac field, and  additionally that the quaternionic Dirac field admits a non-associative algebra that is not observed in the complex theory, in the same token of the scalar field. In principle, physical phenomena that require non-associative symmetry may be described in terms of a quaternionic theory,
and this fact evinces the idea depicted in the first paragraph of this introduction, that the introduction of mathematical structure improves the descriptive capacity of a physical theory. In our particular case, the quaternionic framework enables the quantum theory to describe non-associative physical phenomena that cannot be undertaken over complex bases, and this is evidently an important result.

\section{ The quaternionic Dirac equation\label{D}}
In order to establish the notation, a quaternion $\,q\,$ is a hyper-complex number that can be written as
\begin{equation}\label{d01}
 q=x_0 + x_1 i + x_2 j + x_3 k, 
\end{equation}
where
\begin{equation} x_0,\,x_1,\,x_2,\,x_3\in\mathbbm{R}\qquad\mbox{and}\qquad i^2=j^2=k^2=-1.
\end{equation}	
The imaginary units $\,i,\,j\,$ and $\,k,\,$ represented by $\,e_a,\,$  satisfy the anti-commutative law
\begin{equation}\label{d02}
e_a e_b =\epsilon_{abc}e_c-\delta_{ab},\qquad\mbox{where}\qquad a,\,b,\,c=\big\{1,\,2,\,3\big\},
\end{equation}
$\,\epsilon_{abc}\,$ is the Levi-Civit\`a tensor, and $\,\delta_{ab}\,$ represents the Kronecker delta. The four-dimensional real notation (\ref{d01}) can be turned into the symplectic two-dimensional  complex notation, so that
\begin{equation}\label{d03}
q=z_0+z_1j,\qquad\mbox{where}\qquad z_0=x_0+x_1i\qquad\textrm{and}\qquad z_1=x_2+x_3i.
\end{equation}
Therefore, the quaternionic wave function $\Psi$ in symplectic notation reads
\begin{equation}\label{d04}
\Psi=\cos\Theta\,\psi^{(0)}+\sin\Theta \,\psi^{(1)}\,j,
\end{equation}
where $\,\psi^{(\alpha)}\,$ are complex spinor functions, $\,\alpha=\{0,\,1\},\,$ and $\Theta$ is a real function. In order to define the quaternionic Dirac equation, we recall the generalized quaternionic four-momentum \cite{Giardino:2019xwm,Giardino:2021lov,Giardino:2021mjj}
\begin{equation}\label{d05}
\widehat\Pi^\mu\Psi=\big(\partial^\mu -\mathcal A^\mu\big)\Psi\, i,\qquad\mbox{where}\qquad\mathcal A^\mu=a^\mu i+b^\mu j.
\end{equation}
The quaternionic gauge potential $\mathcal A^\mu$ is pure imaginary, where $a^\mu$ is  real, $b^\mu$ is  complex, and consequently $\,(\mathcal A^\mu)^*=-\mathcal A^\mu.\,$ We also notice that the classical significance of the quaternionic gauge potential has already been explored in \cite{Giardino:2020uab}. Adopting the system of units $\,\hslash=c=1,\,$  the quaternionic Dirac equation for a particle of mass $\,m\,$ reads
\begin{equation}\label{d06}
\left(\widehat{\slashed\Pi} -m\right)\Psi=0,\qquad\mbox{where}\qquad\widehat{\slashed\Pi}=\gamma_\mu \widehat\Pi^\mu,
\end{equation}
and $\gamma^\mu$ represents each $4\times 4$ Dirac matrix. The important detail to be observed in the quaternionic equation is the right hand position of the imaginary unit $\,i\,$ defined in the four-momentum (\ref{d05}). Using the wave function (\ref{d04}) and the equation (\ref{d06}), we have
\begin{eqnarray}
\nonumber &&\cos\Theta\Big(i\slashed\partial+\slashed a -m \Big)\psi^{(0)}-i\sin\Theta\Big(\slashed\partial\Theta\,\psi^{(0)}-\slashed b\,\psi^{*(1)}\Big)+\\
\label{d07}
&&+\Bigg[\sin\Theta\Big(i\slashed\partial+\slashed a+m\Big)\psi^{(1)}+i\cos\Theta\Big(\slashed\partial\Theta\,\psi^{(1)}-\slashed b\,\psi^{*(0)}\Big)
\Bigg]j=0.
\end{eqnarray}
The above equation can be subsumed within complex equations and constraints
\begin{eqnarray}
&&\nonumber \Big(i\slashed\partial+\slashed a -m \Big)\psi^{(0)}=0\\
&&\label{d08}\Big(i\slashed\partial+\slashed a+m\Big)\psi^{(1)}=0\\
&&\nonumber\slashed\partial\Theta\,\psi^{(\alpha)}-\slashed b\,\psi^{*(\alpha')}=0,
\end{eqnarray}
where $\,\alpha\neq \alpha'\,$ and $\,\alpha=\big\{0,\,1\big\}.\,$
In the simplest case, the gauge four-potential is zero, and consequently $\,a^\mu=b^\mu=0.\,$
Considering the wave function
\begin{equation}\label{d09}
\Psi=\cos\Theta\,\exp\Big[ik^{(0)}_\mu x^\mu\Big]u^{(0)}+\sin\Theta\,\exp\Big[ik^{(1)}_\mu x^\mu\Big]u^{(1)}\,j,
\end{equation}
where $u^{(\alpha)}$ are constant spinors, and $\,k^{(\alpha)\mu}=\Big(k^{(\alpha)0},\,\bm k^{(\alpha)}\Big)\,$   are constant four vectors.
Therefore, (\ref{d08}-\ref{d09}) imply that
\begin{equation}\label{d10}
k^{(\alpha)\mu}k^{(\alpha)}_\mu=m^2\qquad\qquad\mbox{and}\qquad\qquad
\partial_\mu\Theta\,\partial^\mu\Theta=0.
\end{equation}
Proposing the transformation
\begin{equation}\label{d11}
\psi^{(\alpha)}\to e^{\pm i\Theta}\psi^{(\alpha)}
\end{equation}
and the function
\begin{equation}\label{d12}
\Theta=\theta_\mu x^\mu+\Theta_0,
\end{equation}
where $\Theta_0$ is a real constant, the complex components of the quaternionic Dirac equation turns into
\begin{equation}
\big(i\slashed\partial-m\big)\psi^{(0)}=0
\qquad\mbox{and}\qquad
\big(i\slashed\partial+m\big)\psi^{(1)}=0.
\end{equation}
The flipped signals indicate that the spinors $u^{(\alpha)}$ of each wave function are not identical. Remembering that the Dirac spinor is composed of two components of two dimensions each, these components occupy changed positions in the spinor. In other words, if
$u^{(0)}=(u,\,v)$, then $u^{(1)}=(v,\,u)$. Finally, defining the momentum
\begin{equation}
p^{(\alpha)\mu}=k^{(\alpha)\mu}\pm\theta^\mu
\end{equation}
we obtain the constraints
\begin{equation}\label{d13}
p^{(\alpha)\mu}p^{(\alpha)}_\mu=m^2,\qquad
\theta^\mu\theta_\mu=0\qquad\mbox{and}\qquad k^{(\alpha)\mu}\theta_\mu=0.
\end{equation}
The situation is thus very similar to that found in the case of the quaternionic scalar field \cite{Giardino:2021lov}, where the quaternionic model turns into two constrained complex models as well. The quantization strategy will accordingly be very similar in both of the cases, where two
quantization schemes emerge. Let us then consider the first of them.

\section{The four components quantization}

Following the quaternionic scalar field approach \cite{Giardino:2022kxk}, let us propose four complex fields
\begin{equation}\label{q01}
\Psi^{(1)}=e^{i\Theta}\psi^{(0)},\qquad 
\Psi^{(2)}=e^{-i\Theta}\psi^{(0)},\qquad
\Psi^{(3)}=e^{i\Theta}\psi^{(1)},\qquad
\Psi^{(4)}=e^{-i\Theta}\psi^{(1)},
\end{equation}
and the quaternionic field accordingly is
\begin{equation}\label{q02}
\Psi=\frac{1}{2}\left(\Psi^{(1)}+\Psi^{(2)}\right)+\frac{1}{2i}\left(\Psi^{(3)}-\Psi^{(4)}\right)j.
\end{equation}
The field is of course constrained, as we see from
\begin{equation}
 \Psi^{(1)}\Psi^{(4)}=\Psi^{(2)}\Psi^{(3)},\qquad\qquad\mbox{and}\qquad\qquad \Psi^{(1)}\Psi^{(3)\dagger}=\Psi^{(2)}\Psi^{(4)\dagger},
\end{equation}
and these constraints are already considered in the momentum constraints (\ref{d13}).
Following the usual complex quantization, the Lagrangian density will be
\begin{equation}\label{q03}
\mathcal L=\sum_{a=1}^4\mathcal L^{(a)},
\end{equation}
where
\begin{equation}
\mathcal L^{(a)}=
\left\{
\begin{array}{l}
\overline{\Psi}^{(a)}\slashed\partial\Psi^{(a)}i-m\,\overline\Psi^{(a)}\Psi^{(a)}\qquad\mbox{for}\qquad a=\big\{1,\,2\big\}\\ 
\\
\overline{\Psi}^{(a)}\slashed\partial\Psi^{(a)}i+m\,\overline\Psi^{(1)}\Psi^{(a)}\qquad
\mbox{for}\qquad a=\big\{3,\,4\big\},
\end{array}
\right.
\end{equation}
and the adjoint wave function is defined in the straightforward way as $\,\overline\Psi^{(a)}=\Psi^{(a)\dagger}\gamma^0.\,$
Subsequently, the Hamiltonian density is
\begin{equation}\label{q04}
\mathcal H=\sum_{a=1}^4\mathcal H^{(a)},
\end{equation}
where
\begin{equation}
\mathcal H^{(a)}=\left\{
\begin{array}{rl}
m\,\overline\Psi^{(a)}\Psi^{(a)}-\overline{\Psi}^{(a)}\bm{\gamma\cdot\nabla}\Psi^{(a)}i & \qquad \mbox{for}\qquad a=\big\{1,\,2\big\}\\
& \\
-m\,\overline\Psi^{(a)}\Psi^{(a)}-\overline{\Psi}^{(a)}\bm{\gamma\cdot\nabla}\Psi^{(a)}i &\qquad\mbox{for}\qquad a=\big\{3,\,4\big\}.
\end{array}
\right.
\end{equation}
Imposing anti-commutation relations to the complex components of the quaternionic Dirac field (\ref{q01}), we get
\begin{equation}\label{q05}
\left\{\widehat\Psi^{(a)}_M(x),\,\widehat\Psi^{(b)\dagger}_N(y)\right\}=\delta^{ab}\delta_{MN}\delta^3(x-y)
\end{equation}
and
\begin{equation}\label{q06}
\left\{\widehat\Psi^{(a)}_M(x),\,\widehat\Psi^{(b)}_N(y)\right\}=\left\{\widehat\Psi^{(a)\dagger}_M(x),\,\widehat\Psi^{(b)\dagger}_N(y)\right\}=0.
\end{equation}
Accordingly, the quantized fields are straightforwardly given  from the usual Dirac quantized field
\begin{equation}\label{q07}
\widehat\Psi_M^{(a)}(x)=\sum_{s=\pm\frac{1}{2}}d^3p^{(a)}\sqrt{\frac{m}{(2\pi)^3p^{(a)0}}}
\Bigg\{\exp\left[-ip^{(a)}_\mu x^\mu\right]\widehat c^{(a)}_s \left(u^{(a)}_s\right)_M
+\exp\left[ip^{(a)}_\mu x^\mu\right]\widehat d^{(a)\dagger}_s \left(v^{(a)}_s\right)_M\Bigg\},
\end{equation}
and the annihilation operators of positive and negative energies are respectively
\begin{equation}
\widehat c^{(a)}_s=\widehat c\left(\bm p^{(a)},\,s\right)\qquad \widehat d^{(a)}_s=\widehat d\left(\bm p^{(a)},\,s\right)
\end{equation}
whose adjoints $\widehat c^{(a)\dagger}_s$ and $\widehat d^{(a)\dagger}_s$ are creation operators of positive and negative energies. Moreover,
\begin{equation}
u^{(a)}_s=u\left(\bm p^{(a)},\,s\right)\qquad v^{(a)}_s=v\left(\bm p^{(a)},\,s\right)
\end{equation}
 are respectively positive and negative energy spinors. The states of positive energy are then
\begin{eqnarray}
\nonumber 
&&c^{(1)\dagger}_s\big|0\big\rangle=\big|\theta,\,k^{(1)},\,s\big\rangle,\qquad
c^{(2)\dagger}_s\big|0\big\rangle=\big|-\theta,\,k^{(1)},\,s\big\rangle,\\
\label{q08}
&&c^{(3)\dagger}_s\big|0\big\rangle=\big|\theta,\,k^{(2)},\,s\big\rangle,\qquad
c^{(4)\dagger}_s\big|0\big\rangle=\big|-\theta,\,k^{(2)},\,s\big\rangle,
\end{eqnarray}
with operator $\widehat d^\dagger$ acting likewise. These commutators satisfy the anti-commuting algebra
\begin{equation}\label{q09}
\left\{\widehat c^{(a)}_{s},\,\widehat c^{\dagger(b)}_{s'}\right\}=
\left\{\widehat d^{(a)}_{s},\,\widehat d^{\dagger(b)}_{s'}\right\}=\delta_{ss'}\delta^{ab}\delta^3(x-y),
\end{equation}
with all other anti-commuting relations vanishing. The above relations come immediately from the usual complex Dirac field, and consequently we can obtain the normal ordered Hamiltonian operator from the Hamiltonian density (\ref{q04}) in terms of annihilation and creation operators
\begin{equation}\label{q10}
:\widehat H:=\sum_{a=1}^4:\widehat H^{(a)}:
\end{equation}
where
\begin{equation}\label{q11}
:\widehat H^{(a)}:\sum_{s=\pm\frac{1}{2}}\int d^3p^{(a)}\, p^{(a)0}
\left[\,\widehat c^{\dagger(a)}_{s}\widehat c^{(a)}_{s}
+\widehat d^{\dagger(a)}_{s}\widehat d^{(a)}_{s}\,\right].
\end{equation}
In the same token, we obtain the charge operator
\begin{equation}\label{q12}
:\widehat Q:=\sum_{a=1}^4:\widehat Q^{(a)}:
\end{equation}
where
\begin{equation}\label{q13}
:\widehat Q^{(a)}:\sum_{s=\pm\frac{1}{2}}\int d^3p^{(a)}\, p^{(a)0}
\left[\,\widehat c^{\dagger(a)}_{s}\widehat c^{(a)}_{s}
-	\widehat d^{\dagger(a)}_{s}\widehat d^{(a)}_{s}\,\right]
\end{equation}
Now we have to connect these results in order to recover the quaternionic wave function. Let us recall the complex result
\begin{equation}\label{q14}
\big\langle \widehat\Psi^{(a)}\big| p^{(b)}\big\rangle=\delta^{ab}\frac{1}{\sqrt{(2\pi)^3 \big|2p^{(a)}_0\big|}}\exp\left[-ip^{(a)}_\mu x^\mu\right]\left(u^{(a)}_s\right)_M
\end{equation}
In order to obtain the wave function, we may define the operator
\begin{equation}\label{q15}
\widehat c=\sum_{a=1}^4\sum_{s_a=\pm 1/2}\sqrt{\big|2p^{(a)}_0\big|}\widehat c^{(a)}_{s_a},\qquad\mbox{and}\qquad
\widehat d=\sum_{a=1}^4\sum_{s_a=\pm 1/2}\sqrt{\big|2p^{(a)}_0\big|}\widehat d^{(a)}_{s_a}.
\end{equation}
We notice that $\,s_1=s_2\,$ and $\,s_3=s_4\,$ in order to satisfy the usual quaternionic Dirac equation in terms of the symplectic wave function (\ref{d04}). By way of example
\begin{equation}
\left\langle 0\big|\,\widehat c\,\big|\widehat\Psi\right\rangle=\frac{1}{(2\pi)^\frac{3}{2}}\Psi.
\end{equation}
Furthermore, the possible quaternionic quantum states are
\begin{equation}
\left|\widehat\Psi\right\rangle=\left\{
\begin{array}{l}
\Big[\frac{1}{2}\left(\widehat\Psi^{(1)\dagger}+\widehat\Psi^{(2)\dagger}\right)+\frac{1}{2i}\left(\widehat\Psi^{(3)\dagger}-\widehat\Psi^{(4)\dagger}\right)j\,\Big]\Big|\,0	\Big\rangle\\ \\
\Big[\frac{1}{2}\left(\widehat\Psi^{(1)}+\widehat\Psi^{(2)}\right)+\frac{1}{2i}\left(\widehat\Psi^{(3)\dagger}-\widehat\Psi^{(4)\dagger}\right)j\,\Big]\Big|\,0	\Big\rangle\\ \\
\Big[\frac{1}{2}\left(\widehat\Psi^{(1)\dagger}+\widehat\Psi^{(2)\dagger}\right)+\frac{1}{2i}\left(\widehat\Psi^{(3)}-\widehat\Psi^{(4)}\right)j\,\Big]\Big|\,0	\Big\rangle\\ \\
\Big[\frac{1}{2}\left(\widehat\Psi^{(1)}+\widehat\Psi^{(2)}\right)+\frac{1}{2i}\left(\widehat\Psi^{(3)}-\widehat\Psi^{(4)}\right)j\,\Big]\Big|\,0	\Big\rangle
\end{array}
\right.
\end{equation}
And the quantization of the field in terms of their four complex components is now complete. The quantization procedure is straightforward after the assumption of the Lagrangian formalism in terms of the complex components, instead of a pure quaternionic formalism that would require the definition of a quaternionic derivative. Before going to the next section, we observe that the quaternionic Dirac field does not allow a nonassociative structure observed in the quaternionic scalar field \cite{Giardino:2021lov}. Although seemingly irrelevant, this difference points to a fundamental difference between of the physical objects that may be described by the scalar and fermionic fields.

\section{The two components quantization\label{T}}
Using the constant function $\Theta=\Theta_0$, the symplectic wave function (\ref{d04}) reads
\begin{equation}\label{t01}
\Psi=\cos\Theta_0\,\Psi^{(0)}+\sin\Theta_0 \,\Psi^{(1)}\,j.
\end{equation}
Following the procedure developed in the four components case, the quantization method is straightforward. We first define the Lagrangian density 
\begin{equation}\label{t02}
\mathcal L=\cos^2\Theta_0\mathcal L^{(0)}+\sin^2\Theta_0 \mathcal L^{(1)},
\end{equation}
where
\begin{equation}
\mathcal L^{(\alpha)}=
\left\{
\begin{array}{l}
\overline{\Psi}^{(0)}\slashed\partial\Psi^{(a)}i-m\,\overline\Psi^{(a)}\Psi^{(0)}\,\\ \\
\overline{\Psi}^{(1)}\slashed\partial\Psi^{(1)}i+m\,\overline\Psi^{(1)}\Psi^{(1)}.
\end{array}
\right.
\end{equation}
In the same manner, the Hamiltonian density is
\begin{equation}\label{t03}
\mathcal H=\cos^2\Theta_0\mathcal H^{(0)}+\sin^2\Theta_0\mathcal H^{(1)},
\end{equation}
and
\begin{equation}
\mathcal H^{(\alpha)}=\left\{
\begin{array}{r}
m\,\overline\Psi^{(0)}\Psi^{(0)}-\overline{\Psi}^{(0)}\bm{\gamma\cdot\nabla}\Psi^{(0)}i\\ \\
-m\,\overline\Psi^{(a)}\Psi^{(1)}-\overline{\Psi}^{(1)}\bm{\gamma\cdot\nabla}\Psi^{(1)}i.
\end{array}
\right.
\end{equation}
The quantized complex components of the Dirac field are
\begin{equation}\label{t04}
\widehat\Psi_M^{(\alpha)}(x)=\sum_{s=\pm\frac{1}{2}}d^3k^{(a)}\sqrt{\frac{m}{(2\pi)^3k^{(a)0}}}
\Bigg\{\exp\left[-ik^{(a)}_\mu x^\mu\right]\widehat c^{(a)}_s \left(u^{(a)}_s\right)_M
+\exp\left[ik^{(a)}_\mu x^\mu\right]\widehat d^{(a)\dagger}_s \left(v^{(a)}_s\right)_M\Bigg\}
\end{equation}
Accordingly, the Hamiltonian operator is
\begin{equation}\label{t05}
:\widehat H:=\cos^2\Theta_0:\widehat H^{(0)}:\,+\,\sin^2\Theta_0 :\widehat H^{(1)}:
\end{equation}
where
\begin{equation}\label{t06}
:\widehat H^{(\alpha)}:\sum_{s=\pm\frac{1}{2}}\int d^3k^{(a)}\, k^{(a)0}
\left[\,\widehat c^{\dagger(a)}_{s}\widehat c^{(a)}_{s}
+\widehat d^{\dagger(a)}_{s}\widehat d^{(a)}_{s}\,\right].
\end{equation}
In the same token, we obtain the charge operator
\begin{equation}\label{t07}
:\widehat Q:=\cos^2\Theta_0:\widehat Q^{(0)}:\,+\,\sin^2\Theta_0 :\widehat Q^{(1)}:
\end{equation}
where
\begin{equation}\label{t08}
:\widehat Q^{(\alpha)}:\sum_{s=\pm\frac{1}{2}}\int d^3k^{(a)}\, k^{(a)0}
\left[\,\widehat c^{\dagger(a)}_{s}\widehat c^{(a)}_{s}
-\widehat d^{\dagger(a)}_{s}\widehat d^{(a)}_{s}\,\right].
\end{equation}
Finally, the possible quantum state operators are as follows
\begin{equation}
\widehat \Psi=\left\{
\begin{array}{l}
\cos\Theta_0\,\Psi^{(0)\dagger}+\sin\Theta_0 \,\Psi^{(1)\dagger}\,j\\
\\
\cos\Theta_0\,\Psi^{(0)\dagger}+\sin\Theta_0 \,\Psi^{(1)}\,j\\
\\
\cos\Theta_0\,\Psi^{(0)}+\sin\Theta_0 \,\Psi^{(1)\dagger}\,j\\
\\
\cos\Theta_0\,\Psi^{(0)}+\sin\Theta_0 \,\Psi^{(1)}\,j.
\end{array}
\right.
\end{equation}

With this result, the correspondence between the quaternionic fermionic field and the quaternionic scalar field is established, and the quantization method seems potent enough  to be further applied to quaternionic fields whose quantization is unknown, like the electromagnetic field.

\section{Conclusion\label{C}}
The quantization of the Dirac field performed in this article confirms the applicability of the method developed for the scalar field \cite{Giardino:2022kxk}, where the
quantized elements are the complex components rather than the complete quaternionic field. This simple proposal avoids the serious difficulty of defining 
quaternionic derivatives to generate the equations of motion from a quaternionic Lagrangian density. 

The quantized Dirac field presents several interesting features. First of all, the higher number of degrees of freedom when compared to the complex field, an expected feature that enables the theory to describe subtler physical phenomena. This expected feature is common to the quaternionic scalar field. However, what is new is that the nonassociative character is not observed in the fermionic field. This reinforces the differences between the scalar field and the fermionic field, and points out that novel and unexpected features may emerge in more sophisticated situations. Therefore, the construction of a quaternionic quantum electrodynamics using the classical quaternionic theory \cite{Giardino:2020uab}  seems a direction of future research where such possibilities may be obtained. 

\paragraph{Data availability statement}The author declares that data sharing is not applicable to this article as no datasets were generated or analysed during the current study.

%
%
%
%
\begin{footnotesize}

\end{footnotesize}
\end{document}